\documentclass[aps,pra,reprint,superscriptaddress,showpacs]{revtex4-1}

\usepackage[utf8]{inputenc}
\usepackage{amsfonts,amsmath,amssymb}
\usepackage{graphicx}
\usepackage{esvect}
\usepackage{units}
\usepackage{paralist}
\usepackage{placeins}
\usepackage{wrapfig}
\usepackage{color}
\usepackage{hyperref}

\begin{document}


\title{Dephasing-Controlled Particle Transport Devices}


\author{Edin Kapetanović}
\email[Contact: ]{ekapetan@itp.uni-bremen.de}
\affiliation{Institut für Theoretische Physik, Universität Bremen, Otto-Hahn-Allee 1, D-28359, Bremen, Germany}

\affiliation{Bremen Center for Computational Materials Science, Universität Bremen, Am Fallturm 1, D-28359, Bremen, Germany}

\author{César A. Rodríguez-Rosario}

\affiliation{Nano-Bio Spectroscopy Group and ETSF, Department of Materials Science, Universidad del País Vasco UPV/EHU, E-20018 San Sebastián, Spain}

\affiliation{Max Planck Institute for the Structure and Dynamics of Matter Hamburg and Center for Free-Electron Laser Science and Department of Physics, University of Hamburg, Luruper Chaussee 149, D-22761, Hamburg, Germany}

\affiliation{Bremen Center for Computational Materials Science, University of Bremen, Am Fallturm 1, D-28359, Bremen, Germany}

\author{Thomas Frauenheim}

\affiliation{Bremen Center for Computational Materials Science, University of Bremen, Am Fallturm 1, D-28359, Bremen, Germany}


\date{January 25, 2017}

\begin{abstract}
We study the role of dephasing in transport through different structures. We show that interference effects invalidate Kirchhoff's circuit laws in quantum devices and illustrate the emergence of ohmic conduction under strong dephasing. We present circuits where the particle transport and the direction of rectification can be controlled through the dephasing strength. This suggests the possibility of constructing molecular devices with new functionalities which use dephasing as a control parameter.
\end{abstract}

\pacs{03.65.Yz, 05.60.Gg}

\maketitle

\section{Introduction}
Electrical devices are reaching the scale where quantum mechanical effects dominate the particle transport. Thus, gaining a deeper understanding of transport processes governed by quantum-mechanical laws is crucial. Since it localizes particles that would otherwise act as delocalized waves, quantum decoherence, also referred to as dephasing, plays an important role. In the last few decades, dephasing has been thoroughly discussed \cite{DecoherenceTransitionQuantumClassicalZurek,DecoherenceMeasurementProblemInterpretation} as it provides an explanation for the transition from quantum to classical. Dephasing in a quantum system is caused by irreversible interactions with an environment and is often seen as undesired noise that destroys desired quantum properties. However, it has also been shown that dephasing can greatly improve the transport efficiency \cite{DephasingEnhancedTransportStrongCorrelation,HeatTransportSpinChain,QuantumExcitationTransportFluctuations,QuantumWalkPhotosynthesis,NoiseAssistedEnergyTransferPhotoSynthesis} by inhibiting destructive interference effects. Symmetries and disorder in networks have also been investigated as key factors for quantum enhanced transport efficiencies \cite{OptimallyDesignedQTransportDisorderedNetworks}. In contrast to those works, we focus on particle currents through networks rather than the transport efficiency of a single excitation.

Through the theory of open quantum systems, we use a master equation approach to model simple quantum transport devices. By numerically solving the master equation for different systems, we determine properties such as the resistance of a set of quantum circuits and compare them to classical expectations. We also investigate the role of dephasing for the particle transport. 

A previous experimental result \cite{OhmsLawAtomicScale} showed that Ohm's law remains valid for a nanowire. In contrast, this work addresses the question whether classical laws survive in more complicated quantum circuits. An early analysis \cite{ConductanceMolWiresParallel} and further simulations and measurements \cite{KirchhoffLawMolecularCircuits} already provided the result that, on the scale of electron wavelengths, interference effects invalidate Kirchhoff's circuit laws for parallel circuits. We show that our simple model reproduces these results and can thus be used to investigate quantum interference effects in complicated geometries. Said parallel circuits are used to exemplify the emergence of Kirchhoff's laws when applying strong dephasing. Also here we demonstrate that resistors in the quantum regime are not additive. Additionally, a device which only conducts under dephasing is presented. Furthermore, we illustrate how, in a triangular circuit with rectification properties, it is possible to control the direction of rectification through controlling the dephasing strength. Recent research \cite{EquivalentResistanceQuantum} hints that complicated networks which exhibit such interesting properties might be realised by simpler systems with equivalent transport properties, leaving a lot of possibilities for future research.

\section{Formalism}
In this work, we describe particle currents as time-continuous quantum stochastic walks \cite{QSwalk} on mathematical graphs that represent circuits. Couplings to external baths are modeled by non-unitary effects. The states $\left\lbrace |i\rangle \right\rbrace$ represent the possible positions on the graph that is interpreted as a circuit. In analogy to a simple tight-binding model, we choose the Laplacian matrix $L$, which contains the information about the connections, to be the quantum system's Hamiltonian $H$:
\begin{align}
	H = L = D - A \label{Hamiltonian}\\
	A_{ij} = \begin{cases} 1 \; \mbox{if vertices}\, i \, \mbox{and} \, j \, \mbox{are connected} \\
	0 \; \mbox{else} \end{cases}\\
	D_{ij} = \delta_{ij} \, d_a
\end{align}
$\delta_{ij}$ is the Kronecker-symbol, while $d_a$ denotes the \textit{degree} of vertex $a$, which is simply the number of vertices connected to it. We describe a quantum system's state through the \textit{density operator} $\rho$ \cite{QuantInformation}: $\rho = \sum_i p_i |\psi_i \rangle \langle \psi_i|$. The evolution of an open quantum system is generally given by a master equation of the form \cite{PositiveDynamicalSemigroups}: $\dot{\rho} = -i \left[ H,\rho \right] + \mathcal{L}(\rho)$.
From here, $\rho$ denotes the density operator's matrix representation in position basis. The first term, which depends on the system's Hamiltonian $H$, causes unitary evolution while $\mathcal{L}(\rho)$ is a superoperator which represents the coupling to external baths. For $\mathcal{L}(\rho) = 0$, this model corresponds to coherent hopping of non-interacting particles between connected sites. The master equation is Markovian if $\mathcal{L}(\rho)$ can be written in \textit{Lindblad form} \cite{OpenQuantumSystemsIntroduction}:
$\mathcal{L}(\rho) = \sum_k \gamma_k \left[ L_k \rho L_k^{\dagger} - \frac{1}{2} \left\lbrace L_k^{\dagger} L_k , \rho \right\rbrace \right]$, $\gamma_k \geq 0$.
$L_k$ denote a complete set of operators and $\left\lbrace \cdot , \cdot \right\rbrace$ is an anti-commutator. The diagonal elements of the density matrix $\rho$ are interpreted as particle populations.

At the boundaries of the system, which are denoted as $|1\rangle$ and $|k\rangle$, particles are being injected and ejected. Explicitly, this injection and ejection is realized by coupling the system incoherently to external baths \cite{ThermoQuantumCoherence, DephasingEnhancedTransportStrongCorrelation, TransportQExcitationsViaFluctuations}, $|L\rangle$ and $|R\rangle$, whose populations are held constant throughout the integration:
\begin{align}
	L_{\text{in}} = \sqrt{\gamma} |1\rangle \langle L| \\
	L_{\text{out}} = \sqrt{\gamma} |R\rangle \langle k|
\end{align}
Setting the rates to $\gamma = 2$ for simplicity, and assuming validity of the Markov approximation, we express the injection and ejection of particles in Lindblad form:
\begin{align}
	\mathcal{L}_L (\rho) = 2 |1\rangle \langle L|\rho |L \rangle \langle 1| - \left\lbrace |L\rangle \langle L| , \rho \right\rbrace \\
	\mathcal{L}_R (\rho) = 2 |R\rangle \langle k|\rho |k \rangle \langle R| - \left\lbrace |R\rangle \langle R| , \rho \right\rbrace
\end{align}
$|1\rangle$ and $|k\rangle$ denote the first and last site in the system. We do not specify whether the system is bosonic or fermionic, which is the reason for choosing the more general first-quantization-expression above. With a constant population $\rho_L = 0.5$ in $|L\rangle$, $\mathcal{L}_L$ causes the injection of 1 particle per arbitrary timestep, which defines the particle current $I = \text{d}N/\text{d}t = 1$.

When electric charge is being transported, the voltage $U$ between two positions is proportional to the difference in charge, or, more accurately, charged particles. In analogy to that, we define the voltage $U$ between two sites on the graph as the difference in population, although it is not necessarily charged particles that are considered. Once the populations become stationary after a sufficiently long time evolution (i.e. the Non-Equilibrium Steady State (NESS) is reached), we determine a resistance $R$ over $R = U / I$. In the systems which are considered, different intial states of $\rho(t = 0)$ are leading to the same NESS, which follows from the known ergodicity of models such as the one considered in this work \cite{ThermalizationErgodicityOpenQSystems}.

Dephasing is a process where quantum coherences are destroyed due to irreversible interactions with an environment. Mathematically, this means that the off-diagonal elements of the density matrix $\rho$ are being destroyed in a certain basis $\left\lbrace |j\rangle \right\rbrace$: $\rho \rightarrow \tilde{\rho} = \sum_j |j\rangle \langle j| \rho |j\rangle \langle j|$. The coupling to a bath which causes dephasing is being modelled by an additional operator $\mathcal{L}_D$:
\begin{align}
\mathcal{L}_D (\rho) = \gamma_D \sum_j [2 |j\rangle \langle j|\rho |j\rangle \langle j| - \left\lbrace |j\rangle \langle j| , \rho \right\rbrace]
\end{align}
with the dephasing rate $\gamma_D \geq 0$. A dephasing rate $\gamma_D$ corresponds to measurements in the chosen basis at a timescale $\approx \gamma_D^{-1}$ \cite{EnvironmentAssistedQTransport}. We define a dimensionless quantity, the dephasing \textit{strength} $\Delta$, as the ratio of the dephasing rate $\gamma_D$ and the rate of coherent hopping $\tau = |A_{\text{ij}}|$ between connected sites, $\Delta = \gamma_D / \tau$. In this work, we focus on the possible positions $\left\lbrace |i\rangle \right\rbrace$ as the basis for dephasing, which implies local interactions with an environment (e.g. surrounding gas molecules randomly hit the system at certain positions). The total master equation for quantum evolution, particle currents and dephasing then reads as:
\begin{align}
\frac{d\rho}{dt} = -i[H,\rho] + \mathcal{L}_L (\rho) + \mathcal{L}_R (\rho) + \mathcal{L}_D (\rho) \label{MasterEquation}
\end{align}
\begin{figure}
	\includegraphics[width=0.5\textwidth]{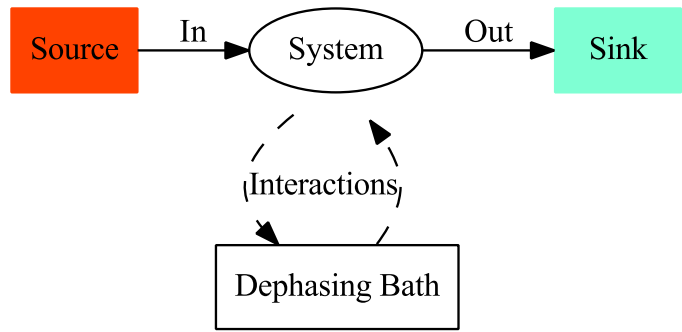}
	\caption{A quantum system is coupled to three Markovian baths, represented by the operators $\mathcal{L}_L$, $\mathcal{L}_R$ and $\mathcal{L}_D$. One bath acts as a particle source, one as a sink, and the third one destroys quantum coherences, which models irreversible interactions with an environment.\label{Graph}}
\end{figure}

We use the concept of the relative entropy \cite{RelativeEntropyVedral} to obtain a gauge for the amount of coherences in a quantum system.
The relative entropy between two density matrices $\rho$ and $\sigma$ is defined as:
\begin{align}
	S(\rho || \sigma) = \text{Tr} (\rho \ln\rho) - \text{Tr} (\rho \ln \sigma )
\end{align}
$S(\rho || \sigma )$ can be interpreted as the amount of information in $\rho$ when assuming the system to be in the state $\sigma$. Here, we choose $\sigma$ to be the coherence-free counterpart of $\rho$: $\sigma = \sum_i |i\rangle \langle i|\rho |i\rangle \langle i|$. The relative entropy $S(\rho || \sigma)$ can then \cite{ThermoQuantumCoherence} be used as a gauge for the amount of coherences with respect to the total particle number in a system.
\section{Results and Discussion}
We now investigate different quantum circuits by defining Hamiltonians $H$ according to their connections (see Eqn.(\ref{Hamiltonian})) and solving Eqn.(\ref{MasterEquation}) \cite{FortranCode}. The most interesting results are presented in this section. Parallel circuits are considered in order to answer the question if classical rules for the resistance are valid. Another simple circuit shows that resistors are not additive in our quantum transport device, which exemplifies violations of Ohmic scaling of the resistance in quantum systems. Additionally, a simple, pentagonal circuit only conducts under dephasing, which allows a controlled particle transport parametrized by dephasing. Furthermore, through control of the dephasing strength, it is possible to control the direction of rectification in a triangular structure.

\subsection{Parallel Circuits}
\begin{figure*}
	\includegraphics[width=1.0\textwidth]{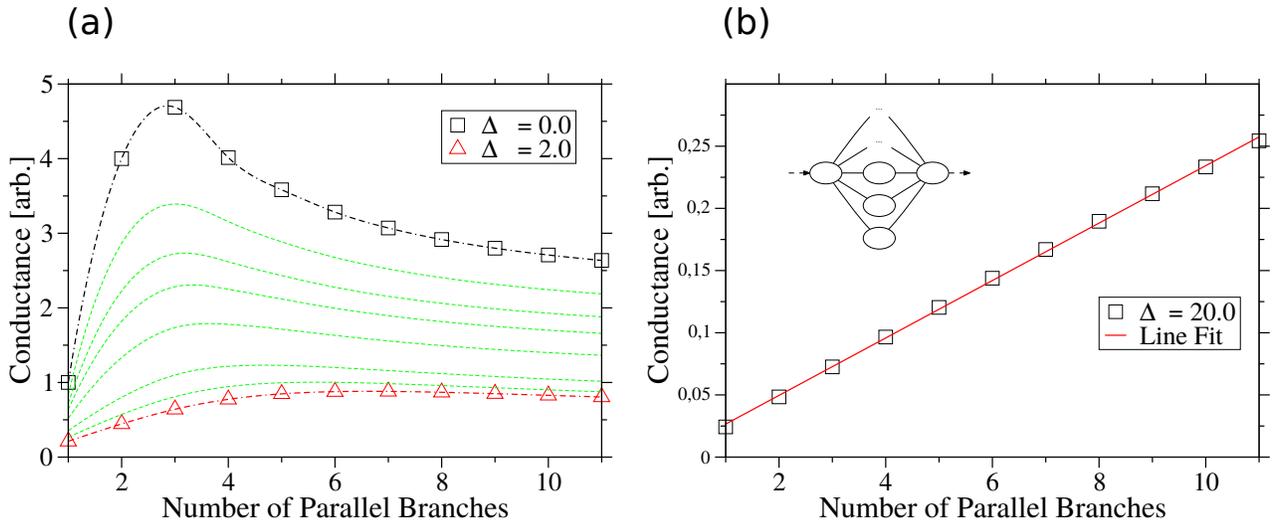}
	\caption{(a): Conductance $G$ vs. number of branches in parallel for different dephasing strengths $\Delta$. In absence of dephasing, using two parallel branches instead of a single wire quadruples the conductance in the quantum case, which is in accordance with other results where actual molecules were discussed. A third parallel branch only increases the conductance slightly, additional branches reduce the conductance due to destructive interference. Applying dephasing flattens the curve. (b): Same curve as in (a) for $\Delta = 20.0$. It illustrates how linear scaling, which is expected according to Kirchhoff's circuit laws, emerges under strong dephasing. The graph in the top left corner illustrates the circuits. The ovals are the nodes which represent possible positions, the edges are interpreted as resistors.\label{Para}}
\end{figure*}
As a starting point, we put classical rules, which remain valid for a nanowire \cite{OhmsLawAtomicScale}, to the test for a set of simple quantum circuits. We determine the effective resistances of parallel circuits with varying number of branches (see the top left corner of Fig.(\ref{Para}b)). In Fig.(\ref{Para}), the conductance $G$ (which is determined over $G = 1/R$), is plotted in dependence of the number of branches with different dephasing strengths $\Delta$. According to Kirchhoff's circuit laws, the conductance scales linearly with the number of branches, which is violated in quantum systems. The black curve in Fig.(\ref{Para}a) shows the behaviour in absence of dephasing. Using two branches instead of a single wire quadruples the conductance due to constructive interference between the particles. Under idealized conditions, a Green's function approach for molecular circuits consisting of two parallel branches yielded a similar result \cite{KirchhoffLawMolecularCircuits}. Here, we also show that a third branch only slightly increases the conductance, while further branches reduce the conductance due to destructive interference effects. When applying dephasing, the curve flattens until (see Fig.(\ref{Para})b) linear scaling, which corresponds to classical behaviour, is reproduced. It should also be noted that, while the curve flattens when increasing the dephasing strength, the peak position moves. The number of branches which maximizes the conductance can be accurately estimated by rounding the linear expression $P(\Delta) = 2.785 + 1.909\cdot \Delta$ to the closest integer. For example, when the dephasing rate is comparable to the system's dynamics (i.e. $\Delta \approx 1$), the conductance peaks when using 5 parallel branches. However, this relation is relevant only for small values of $\Delta$ as the peak vanishes for $\Delta \gg 1$. For the transport of a single excitation, it was already shown \cite{OptimizationExcitonTrapping} that, under strong dephasing, such networks can be accurately mapped to localized, kinetic systems with quantum corrections. Although the models differ (i.e. constant in- and outflux of particles vs. trapping of a single excitation), this helps to understand how classical behaviour emerges. The reference also discusses a two-branch system similar to ours and also demonstrates the crucial role of quantum interference.

\subsection{Additivity}
\begin{figure*}
	\includegraphics[width=1.0\textwidth]{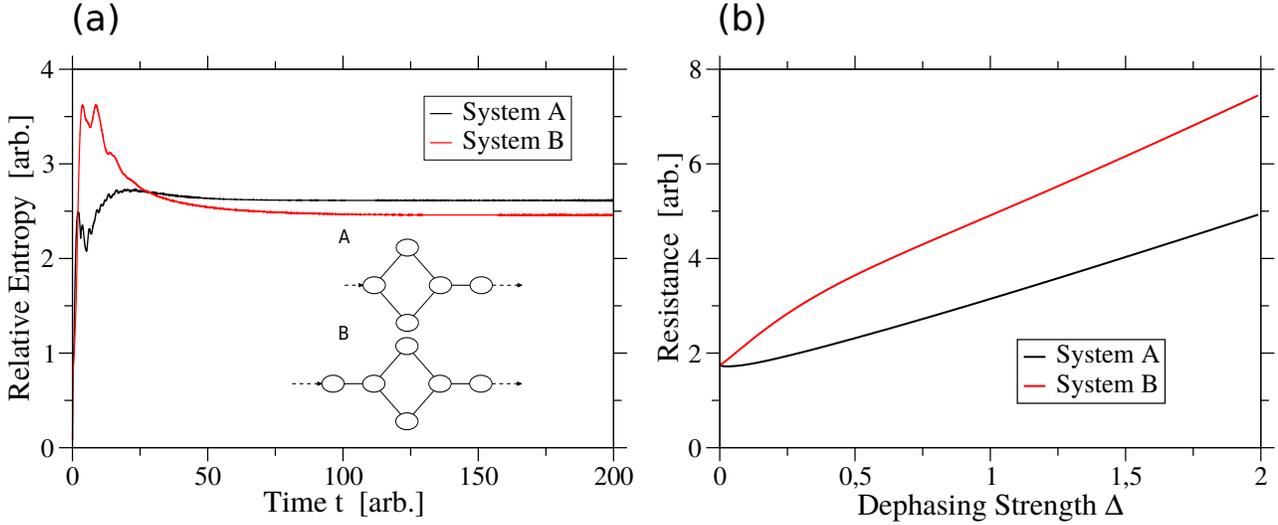}
	\caption{(a): Relative Entropy over time for the systems illustrated by the graphs. In absence of dephasing, both systems have the same resistance although system B has an additional resistor on top. When the NESS is reached, system B has a lower relative entropy, which implies less coherences and thus less destructive interference effects blocking the way. This explains the more efficient particle transport through B. (b): Resistance vs. Dephasing Strength $\Delta$ for both systems. Under dephasing, system B's resistance is higher than A's, which is expected classically.\label{Additivity}}
\end{figure*}

The next question we adress is whether quantum resistors are additive. We send a current through the two systems which are illustrated by the graphs in Fig.(\ref{Additivity}a). For both systems, the determined effective resistance is $R = 1.75$ without dephasing. This means that adding a single resistor on top of system A does not affect the resistance, which shows that, for non-trivial circuits, quantum resistors are not additive. Thus, Ohmic Scaling of the resistance is not generally valid in quantum systems. The fact that the resistances are the same implies that the particle transport from one end to the other is more efficient in system B. In order to understand this phenomenon, we consider the relative entropy between both system's density matrices and their decohered counterparts, as seen in Fig.(\ref{Additivity}a). As mentioned before, the relative entropy between a density matrix and its decohered counterpart can be used as a gauge for coherence with respect to the particle number. System B has a lower relative entropy at the NESS, which implies less coherences and thus, less destructive interference effects blocking the way. Under dephasing (see Fig.(\ref{Additivity}b)), this effect gets diminished and, as expected classically, the bigger system has a higher resistance.

\subsection{Pentagonal Circuit}
\begin{figure}
	\includegraphics[width=0.5\textwidth]{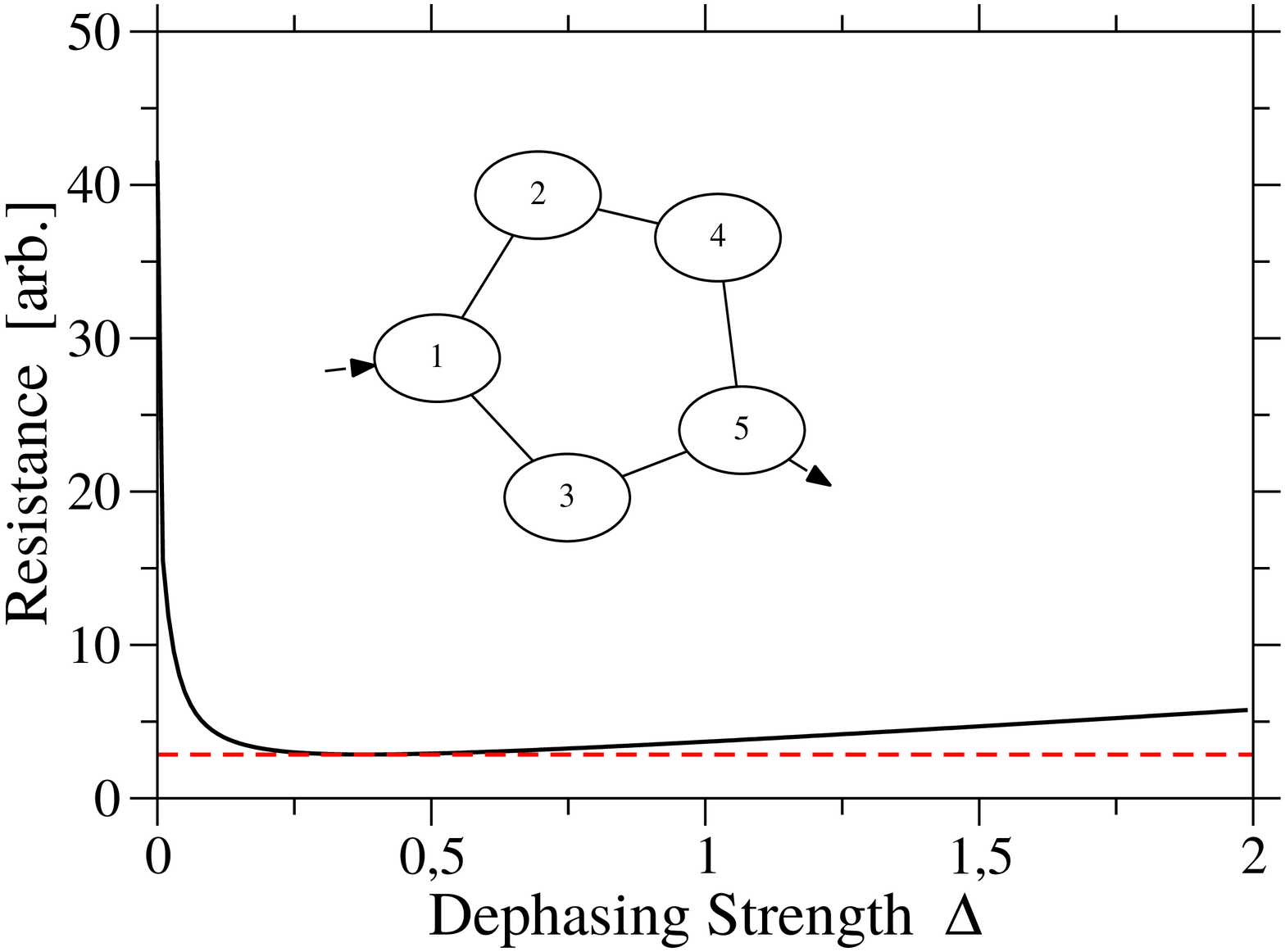}
	\caption{Illustration of the 5-level system and its Resistance depending on the dephasing strength $\Delta$. For $\Delta = 0$, the resistance is infinitely high due to strong destructive interference. When introducing weak dephasing, the resistance sharply falls down to its minimum and then, under stronger dephasing, increases again due to the Zeno effect. Optimal particle transport takes place in a regime between quantum and classical.\label{Penta}}
\end{figure}
We consider a 5-level-system which demonstrates how strong destructive interference can completely block a path. Fig.(\ref{Penta}) shows the pentagonal system and the resistance vs. the dephasing strength. For $\Delta = 0$, the particle distribution does not reach a steady state, which means that the system cannot be seen as a conductor. Comparing the system to the two-branch parallel circuit illustrates how strongly the conductance depends on molecular symmetries. Along with the role of local dephasing, this was also emphasized by other authors \cite{DynamicalSignaturesMolecularSymmetries}. In accordance to our findings, meta-benzene, which resembles our 5-level-system, exhibits a drastically lower conductance \cite{CoherentMolecularConduction} than para-benzene, which has a symmetry similar to the two-branch parallel circuit. Destructive interference prevents most particles from reaching the end of the circuit. When introducing weak dephasing, the resistance sharply falls down to its minimum. Afterwards, it increases again due to the quantum Zeno effect \cite{ZenoParadoxQuantumTheory}. This illustrates the fact \cite{QuantumWalkPhotosynthesis} \cite{DephasingEnhancedTransportStrongCorrelation} that, in some systems, optimal particle transport occurs in a regime between quantum and classical. Through controlling the dephasing strength, it can be controlled whether the circuit acts as an insulator or a conductor.

\subsection{Triangular Circuits}
\begin{figure*}
	\includegraphics[width=1.0\textwidth]{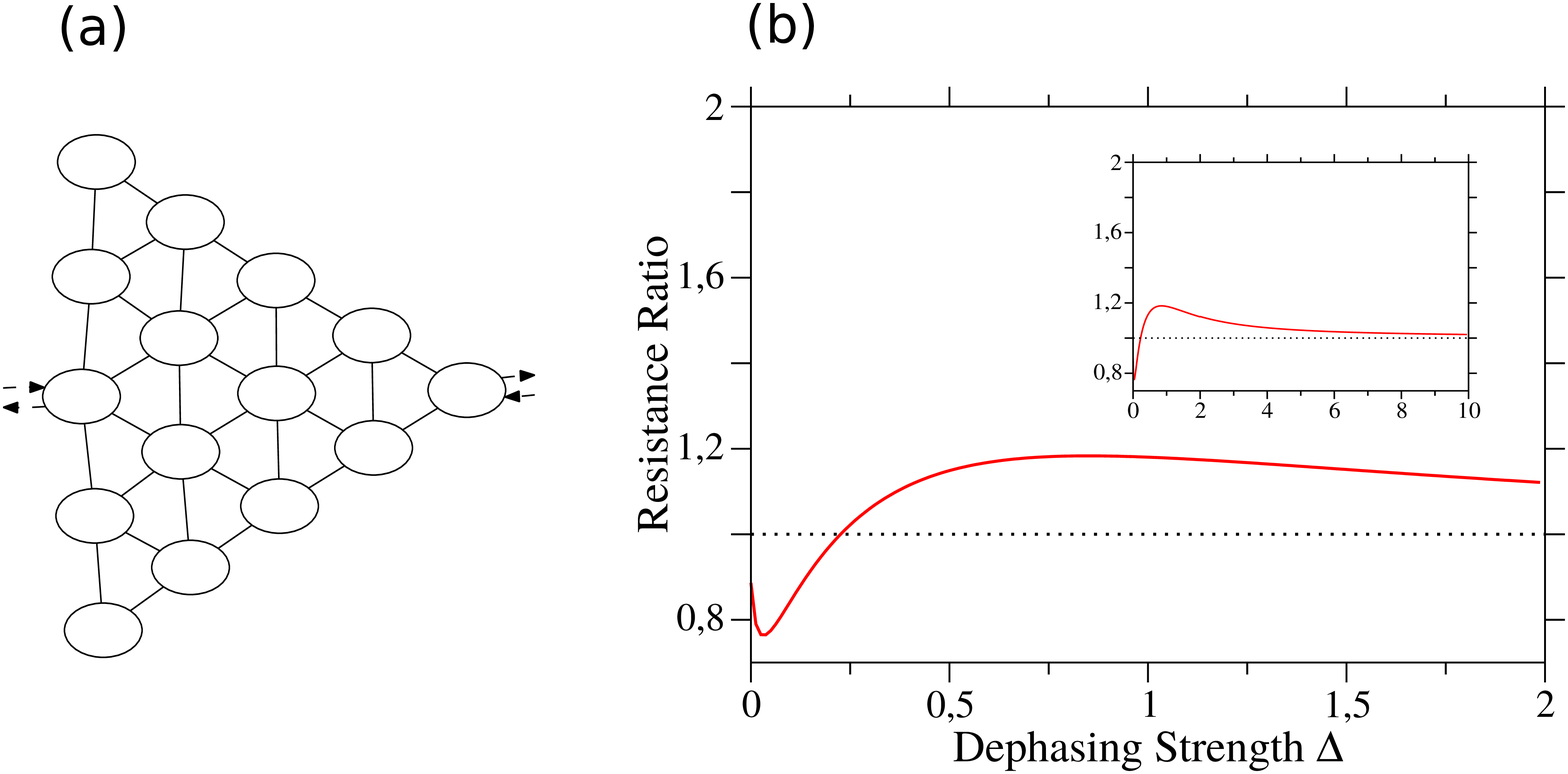}
	\caption{(a): Triangular structure with rectification properties. A current is sent in both directions through the circuit, and the resistances are compared in order to estimate the strength of the rectification. (b): Resistance Ratio vs. Dephasing Strength. Depending on the Dephasing Strength $\Delta$, the Resistance Ratio is either below or above 1, which illustrates how the direction of the rectification can be controlled through dephasing. The small picture in the corner shows the same graph for higher $\Delta$'s. The Resistance Ratio converges to 1 under strong dephasing, which corresponds to classical behaviour.\label{TriangleDephasing}}
\end{figure*}

Since it provides spatial assymmetries and resembles a funnel, we consider a triangular structure (see Fig.(\ref{TriangleDephasing}a)) in this section. Triangular electron cavities were shown to possess interesting transport properties \cite{QuantumRatchets}. As single molecules are known to possess rectification properties \cite{UnimolecularRectifiers} when used for conduction, we investigate the rectification properties of the triangle. Explicitly, this means that we determine the resistance for both directions of the current. In experiments \cite{SingleMoleculeDiodesHighRectificationEnvironmentalControl}, a voltage $U$ is usually set and a rectification ratio, $\frac{I_{\rightarrow}}{I_{\leftarrow}}$, measured and interpreted as the strength of rectification. Here, we consider the resistance ratio $\frac{R_{\rightarrow}}{R_{\leftarrow}}$ instead as it is the current $I$ that is being set. Fig.(\ref{TriangleDephasing}b) shows the resistance ratio depending on the dephasing strength. Depending on the dephasing, the resistance ratio is either above or below one, with a point at $\Delta = 0.2259$ where the resistances for both directions are the same. Under strong dephasing, the ratio converges to $1$, which corresponds to classical behaviour of a resistor. As experiments with molecules have already shown rectification ratios of 200 and higher \cite{SingleMoleculeDiodesHighRectificationEnvironmentalControl}, the magnitude of rectification here seems comparably low. However, it is a new result that the direction of rectification can be controlled through controlling the dephasing strength. 
\section{Conclusions}
We devised a model for a simple quantum transport device which allows the study of interference effects in quantum circuits. We showed how interference effects lead to violations of Kirchhoff's circuit laws and how classical expectations emerge when applying strong, local dephasing. We also presented circuits which demonstrate that quantum resistors are not additive in general. Additionally, we discussed the role of quantum coherences for the particle transport by controlling the dephasing strength. Through discussing a specific 5-level-system, we demonstrated that, in certain systems, the most efficient particle transport takes place in the regime between quantum and classical. Furthermore, we investigated rectification properties in a triangular structure and showed that the direction of rectification can be controlled in a regime with weak dephasing. Although dephasing in general seems to be well understood after decades of research, this shows that even seemingly trivial transport models can exhibit yet unknown properties. It also suggests the construction of new, molecular devices which use dephasing as a control parameter.

Future work includes trying to find similar, useful properties in more realistic geometries while additionally considering interactions between the transported particles.
\bibliography{references.bib}

\begin{thebibliography}{29}%
\makeatletter
\providecommand \@ifxundefined [1]{%
 \@ifx{#1\undefined}
}%
\providecommand \@ifnum [1]{%
 \ifnum #1\expandafter \@firstoftwo
 \else \expandafter \@secondoftwo
 \fi
}%
\providecommand \@ifx [1]{%
 \ifx #1\expandafter \@firstoftwo
 \else \expandafter \@secondoftwo
 \fi
}%
\providecommand \natexlab [1]{#1}%
\providecommand \enquote  [1]{``#1''}%
\providecommand \bibnamefont  [1]{#1}%
\providecommand \bibfnamefont [1]{#1}%
\providecommand \citenamefont [1]{#1}%
\providecommand \href@noop [0]{\@secondoftwo}%
\providecommand \href [0]{\begingroup \@sanitize@url \@href}%
\providecommand \@href[1]{\@@startlink{#1}\@@href}%
\providecommand \@@href[1]{\endgroup#1\@@endlink}%
\providecommand \@sanitize@url [0]{\catcode `\\12\catcode `\$12\catcode
  `\&12\catcode `\#12\catcode `\^12\catcode `\_12\catcode `\%12\relax}%
\providecommand \@@startlink[1]{}%
\providecommand \@@endlink[0]{}%
\providecommand \url  [0]{\begingroup\@sanitize@url \@url }%
\providecommand \@url [1]{\endgroup\@href {#1}{\urlprefix }}%
\providecommand \urlprefix  [0]{URL }%
\providecommand \Eprint [0]{\href }%
\providecommand \doibase [0]{http://dx.doi.org/}%
\providecommand \selectlanguage [0]{\@gobble}%
\providecommand \bibinfo  [0]{\@secondoftwo}%
\providecommand \bibfield  [0]{\@secondoftwo}%
\providecommand \translation [1]{[#1]}%
\providecommand \BibitemOpen [0]{}%
\providecommand \bibitemStop [0]{}%
\providecommand \bibitemNoStop [0]{.\EOS\space}%
\providecommand \EOS [0]{\spacefactor3000\relax}%
\providecommand \BibitemShut  [1]{\csname bibitem#1\endcsname}%
\let\auto@bib@innerbib\@empty
\bibitem [{\citenamefont
  {Zurek}(2002)}]{DecoherenceTransitionQuantumClassicalZurek}%
  \BibitemOpen
  \bibfield  {author} {\bibinfo {author} {\bibfnamefont {W.~H.}\ \bibnamefont
  {Zurek}},\ }\href@noop {} {\bibfield  {journal} {\bibinfo  {journal} {Los
  Alamos Science}\ }\textbf {\bibinfo {volume} {27}},\ \bibinfo {pages} {86}
  (\bibinfo {year} {2002})}\BibitemShut {NoStop}%
\bibitem [{\citenamefont
  {Schlosshauer}(2004)}]{DecoherenceMeasurementProblemInterpretation}%
  \BibitemOpen
  \bibfield  {author} {\bibinfo {author} {\bibfnamefont {M.}~\bibnamefont
  {Schlosshauer}},\ }\href@noop {} {\bibfield  {journal} {\bibinfo  {journal}
  {Reviews of Modern Physics}\ }\textbf {\bibinfo {volume} {76}},\ \bibinfo
  {pages} {1267} (\bibinfo {year} {2004})}\BibitemShut {NoStop}%
\bibitem [{\citenamefont {Mendoza-Arenas}\ \emph
  {et~al.}(2013{\natexlab{a}})\citenamefont {Mendoza-Arenas}, \citenamefont
  {Grujic}, \citenamefont {Jaksch},\ and\ \citenamefont
  {Clark}}]{DephasingEnhancedTransportStrongCorrelation}%
  \BibitemOpen
  \bibfield  {author} {\bibinfo {author} {\bibfnamefont {J.~J.}\ \bibnamefont
  {Mendoza-Arenas}}, \bibinfo {author} {\bibfnamefont {T.}~\bibnamefont
  {Grujic}}, \bibinfo {author} {\bibfnamefont {D.}~\bibnamefont {Jaksch}}, \
  and\ \bibinfo {author} {\bibfnamefont {S.~R.}\ \bibnamefont {Clark}},\
  }\href@noop {} {\bibfield  {journal} {\bibinfo  {journal} {Phys. Rev. B}\
  }\textbf {\bibinfo {volume} {87}} (\bibinfo {year} {2013}{\natexlab{a}})},\
  \bibinfo {note} {235130}\BibitemShut {NoStop}%
\bibitem [{\citenamefont {Mendoza-Arenas}\ \emph
  {et~al.}(2013{\natexlab{b}})\citenamefont {Mendoza-Arenas}, \citenamefont
  {Al-Assam}, \citenamefont {Clark},\ and\ \citenamefont
  {Jaksch}}]{HeatTransportSpinChain}%
  \BibitemOpen
  \bibfield  {author} {\bibinfo {author} {\bibfnamefont {J.~J.}\ \bibnamefont
  {Mendoza-Arenas}}, \bibinfo {author} {\bibfnamefont {S.}~\bibnamefont
  {Al-Assam}}, \bibinfo {author} {\bibfnamefont {S.~R.}\ \bibnamefont {Clark}},
  \ and\ \bibinfo {author} {\bibfnamefont {D.}~\bibnamefont {Jaksch}},\
  }\href@noop {} {\bibfield  {journal} {\bibinfo  {journal} {Journal of
  Statistical Mechanics: Theory and Experiment}\ }\textbf {\bibinfo {volume}
  {2013}} (\bibinfo {year} {2013}{\natexlab{b}})}\BibitemShut {NoStop}%
\bibitem [{\citenamefont {Zhang}\ \emph
  {et~al.}(2015{\natexlab{a}})\citenamefont {Zhang}, \citenamefont {Lee},\ and\
  \citenamefont {Sadeghpour}}]{QuantumExcitationTransportFluctuations}%
  \BibitemOpen
  \bibfield  {author} {\bibinfo {author} {\bibfnamefont {M.}~\bibnamefont
  {Zhang}}, \bibinfo {author} {\bibfnamefont {T.~E.}\ \bibnamefont {Lee}}, \
  and\ \bibinfo {author} {\bibfnamefont {H.~R.}\ \bibnamefont {Sadeghpour}},\
  }\href@noop {} {\bibfield  {journal} {\bibinfo  {journal} {Phys. Rev. A}\
  }\textbf {\bibinfo {volume} {91}} (\bibinfo {year} {2015}{\natexlab{a}})},\
  \bibinfo {note} {052101}\BibitemShut {NoStop}%
\bibitem [{\citenamefont {Mohseni}\ \emph {et~al.}(2008)\citenamefont
  {Mohseni}, \citenamefont {Rebentrost}, \citenamefont {Lloyd},\ and\
  \citenamefont {Aspuru-Guzik}}]{QuantumWalkPhotosynthesis}%
  \BibitemOpen
  \bibfield  {author} {\bibinfo {author} {\bibfnamefont {M.}~\bibnamefont
  {Mohseni}}, \bibinfo {author} {\bibfnamefont {P.}~\bibnamefont {Rebentrost}},
  \bibinfo {author} {\bibfnamefont {S.}~\bibnamefont {Lloyd}}, \ and\ \bibinfo
  {author} {\bibfnamefont {A.}~\bibnamefont {Aspuru-Guzik}},\ }\href@noop {}
  {\bibfield  {journal} {\bibinfo  {journal} {J. Chem. Phys.}\ }\textbf
  {\bibinfo {volume} {129}} (\bibinfo {year} {2008})},\ \bibinfo {note}
  {174106}\BibitemShut {NoStop}%
\bibitem [{\citenamefont {Chin}\ \emph {et~al.}(2010)\citenamefont {Chin},
  \citenamefont {Datta}, \citenamefont {Caruso}, \citenamefont {Huelga},\ and\
  \citenamefont {Plenio}}]{NoiseAssistedEnergyTransferPhotoSynthesis}%
  \BibitemOpen
  \bibfield  {author} {\bibinfo {author} {\bibfnamefont {A.~W.}\ \bibnamefont
  {Chin}}, \bibinfo {author} {\bibfnamefont {A.}~\bibnamefont {Datta}},
  \bibinfo {author} {\bibfnamefont {F.}~\bibnamefont {Caruso}}, \bibinfo
  {author} {\bibfnamefont {S.~F.}\ \bibnamefont {Huelga}}, \ and\ \bibinfo
  {author} {\bibfnamefont {M.~B.}\ \bibnamefont {Plenio}},\ }\href@noop {}
  {\bibfield  {journal} {\bibinfo  {journal} {New Journal of Physics}\ }\textbf
  {\bibinfo {volume} {12}} (\bibinfo {year} {2010})}\BibitemShut {NoStop}%
\bibitem [{\citenamefont {Walschaers}\ \emph {et~al.}(2013)\citenamefont
  {Walschaers}, \citenamefont {de~Cossio~Diaz}, \citenamefont {Mulet},\ and\
  \citenamefont {Buchleitner}}]{OptimallyDesignedQTransportDisorderedNetworks}%
  \BibitemOpen
  \bibfield  {author} {\bibinfo {author} {\bibfnamefont {M.}~\bibnamefont
  {Walschaers}}, \bibinfo {author} {\bibfnamefont {J.~F.}\ \bibnamefont
  {de~Cossio~Diaz}}, \bibinfo {author} {\bibfnamefont {R.}~\bibnamefont
  {Mulet}}, \ and\ \bibinfo {author} {\bibfnamefont {A.}~\bibnamefont
  {Buchleitner}},\ }\href@noop {} {\bibfield  {journal} {\bibinfo  {journal}
  {Phys. Rev. Lett.}\ }\textbf {\bibinfo {volume} {111}} (\bibinfo {year}
  {2013})}\BibitemShut {NoStop}%
\bibitem [{\citenamefont {Weber}\ \emph {et~al.}(2012)\citenamefont {Weber}
  \emph {et~al.}}]{OhmsLawAtomicScale}%
  \BibitemOpen
  \bibfield  {author} {\bibinfo {author} {\bibfnamefont {B.}~\bibnamefont
  {Weber}} \emph {et~al.},\ }\href@noop {} {\bibfield  {journal} {\bibinfo
  {journal} {Science}\ }\textbf {\bibinfo {volume} {335}} (\bibinfo {year}
  {2012})},\ \bibinfo {note} {64}\BibitemShut {NoStop}%
\bibitem [{\citenamefont {Magoga}\ and\ \citenamefont
  {Joachim}(1999)}]{ConductanceMolWiresParallel}%
  \BibitemOpen
  \bibfield  {author} {\bibinfo {author} {\bibfnamefont {M.}~\bibnamefont
  {Magoga}}\ and\ \bibinfo {author} {\bibfnamefont {C.}~\bibnamefont
  {Joachim}},\ }\href@noop {} {\bibfield  {journal} {\bibinfo  {journal} {Phys.
  Rev. B}\ }\textbf {\bibinfo {volume} {59}} (\bibinfo {year} {1999})},\
  \bibinfo {note} {16011}\BibitemShut {NoStop}%
\bibitem [{\citenamefont {Vazquez}\ \emph {et~al.}(2012)\citenamefont
  {Vazquez}, \citenamefont {Skouta}, \citenamefont {Schneebeli}, \citenamefont
  {Kamenetska}, \citenamefont {Breslow}, \citenamefont {Venkataraman},\ and\
  \citenamefont {Hybertsen}}]{KirchhoffLawMolecularCircuits}%
  \BibitemOpen
  \bibfield  {author} {\bibinfo {author} {\bibfnamefont {H.}~\bibnamefont
  {Vazquez}}, \bibinfo {author} {\bibfnamefont {R.}~\bibnamefont {Skouta}},
  \bibinfo {author} {\bibfnamefont {S.}~\bibnamefont {Schneebeli}}, \bibinfo
  {author} {\bibfnamefont {M.}~\bibnamefont {Kamenetska}}, \bibinfo {author}
  {\bibfnamefont {R.}~\bibnamefont {Breslow}}, \bibinfo {author} {\bibfnamefont
  {L.}~\bibnamefont {Venkataraman}}, \ and\ \bibinfo {author} {\bibfnamefont
  {M.}~\bibnamefont {Hybertsen}},\ }\href@noop {} {\bibfield  {journal}
  {\bibinfo  {journal} {Nature Nanotechnology}\ }\textbf {\bibinfo {volume}
  {7}},\ \bibinfo {pages} {663} (\bibinfo {year} {2012})}\BibitemShut {NoStop}%
\bibitem [{\citenamefont {Sarkar}\ \emph {et~al.}(2016)\citenamefont {Sarkar},
  \citenamefont {Kröber},\ and\ \citenamefont
  {Morr}}]{EquivalentResistanceQuantum}%
  \BibitemOpen
  \bibfield  {author} {\bibinfo {author} {\bibfnamefont {S.}~\bibnamefont
  {Sarkar}}, \bibinfo {author} {\bibfnamefont {D.}~\bibnamefont {Kröber}}, \
  and\ \bibinfo {author} {\bibfnamefont {D.~K.}\ \bibnamefont {Morr}},\
  }\href@noop {} {\bibfield  {journal} {\bibinfo  {journal} {Phys. Rev. Lett.}\
  }\textbf {\bibinfo {volume} {117}} (\bibinfo {year} {2016})}\BibitemShut
  {NoStop}%
\bibitem [{\citenamefont {Rodríguez-Rosario}\ \emph
  {et~al.}(2010)\citenamefont {Rodríguez-Rosario}, \citenamefont {Whitfield},\
  and\ \citenamefont {Aspuru-Guzik}}]{QSwalk}%
  \BibitemOpen
  \bibfield  {author} {\bibinfo {author} {\bibfnamefont {C.~A.}\ \bibnamefont
  {Rodríguez-Rosario}}, \bibinfo {author} {\bibfnamefont {J.~D.}\ \bibnamefont
  {Whitfield}}, \ and\ \bibinfo {author} {\bibfnamefont {A.}~\bibnamefont
  {Aspuru-Guzik}},\ }\href@noop {} {\bibfield  {journal} {\bibinfo  {journal}
  {Phys. Rev. A.}\ }\textbf {\bibinfo {volume} {81}} (\bibinfo {year}
  {2010})},\ \bibinfo {note} {022323}\BibitemShut {NoStop}%
\bibitem [{\citenamefont {Nielsen}\ and\ \citenamefont
  {Chuang}(2000)}]{QuantInformation}%
  \BibitemOpen
  \bibfield  {author} {\bibinfo {author} {\bibfnamefont {M.~A.}\ \bibnamefont
  {Nielsen}}\ and\ \bibinfo {author} {\bibfnamefont {I.~L.}\ \bibnamefont
  {Chuang}},\ }\href@noop {} {\emph {\bibinfo {title} {Quantum Computation and
  Quantum Information}}}\ (\bibinfo  {publisher} {Cambridge University Press},\
  \bibinfo {address} {Cambridge, UK},\ \bibinfo {year} {2000})\BibitemShut
  {NoStop}%
\bibitem [{\citenamefont {Gorini}\ \emph {et~al.}(1996)\citenamefont {Gorini},
  \citenamefont {Kossakowski},\ and\ \citenamefont
  {Sudarshan}}]{PositiveDynamicalSemigroups}%
  \BibitemOpen
  \bibfield  {author} {\bibinfo {author} {\bibfnamefont {V.}~\bibnamefont
  {Gorini}}, \bibinfo {author} {\bibfnamefont {A.}~\bibnamefont {Kossakowski}},
  \ and\ \bibinfo {author} {\bibfnamefont {E.~C.~G.}\ \bibnamefont
  {Sudarshan}},\ }\href@noop {} {\bibfield  {journal} {\bibinfo  {journal}
  {Journal of Mathematical Physics}\ }\textbf {\bibinfo {volume} {17}}
  (\bibinfo {year} {1996})},\ \bibinfo {note} {821}\BibitemShut {NoStop}%
\bibitem [{\citenamefont {Ángel Rivas}\ and\ \citenamefont
  {Huelga}(2012)}]{OpenQuantumSystemsIntroduction}%
  \BibitemOpen
  \bibfield  {author} {\bibinfo {author} {\bibnamefont {Ángel Rivas}}\ and\
  \bibinfo {author} {\bibfnamefont {S.~F.}\ \bibnamefont {Huelga}},\
  }\href@noop {} {\emph {\bibinfo {title} {Open Quantum Systems. An
  Introduction}}}\ (\bibinfo  {publisher} {Springer},\ \bibinfo {address}
  {Heidelberg},\ \bibinfo {year} {2012})\BibitemShut {NoStop}%
\bibitem [{\citenamefont {Rodríguez-Rosario}\ \emph
  {et~al.}(2013)\citenamefont {Rodríguez-Rosario}, \citenamefont
  {Frauenheim},\ and\ \citenamefont {Aspuru-Guzik}}]{ThermoQuantumCoherence}%
  \BibitemOpen
  \bibfield  {author} {\bibinfo {author} {\bibfnamefont {C.~A.}\ \bibnamefont
  {Rodríguez-Rosario}}, \bibinfo {author} {\bibfnamefont {T.}~\bibnamefont
  {Frauenheim}}, \ and\ \bibinfo {author} {\bibfnamefont {A.}~\bibnamefont
  {Aspuru-Guzik}},\ }\href@noop {} {\  (\bibinfo {year} {2013})},\ \Eprint
  {http://arxiv.org/abs/1308.1245v1} {arXiv:1308.1245v1 [quant-ph]}
  \BibitemShut {NoStop}%
\bibitem [{\citenamefont {Zhang}\ \emph
  {et~al.}(2015{\natexlab{b}})\citenamefont {Zhang}, \citenamefont {Lee},\ and\
  \citenamefont {Sadeghpour}}]{TransportQExcitationsViaFluctuations}%
  \BibitemOpen
  \bibfield  {author} {\bibinfo {author} {\bibfnamefont {M.}~\bibnamefont
  {Zhang}}, \bibinfo {author} {\bibfnamefont {T.~E.}\ \bibnamefont {Lee}}, \
  and\ \bibinfo {author} {\bibfnamefont {H.~R.}\ \bibnamefont {Sadeghpour}},\
  }\href@noop {} {\bibfield  {journal} {\bibinfo  {journal} {Phys. Rev. A}\
  }\textbf {\bibinfo {volume} {91}} (\bibinfo {year}
  {2015}{\natexlab{b}})}\BibitemShut {NoStop}%
\bibitem [{\citenamefont {Žnidarič}\ \emph {et~al.}(2010)\citenamefont
  {Žnidarič}, \citenamefont {Prosen}, \citenamefont {Benenti}, \citenamefont
  {Casati},\ and\ \citenamefont
  {Rossini}}]{ThermalizationErgodicityOpenQSystems}%
  \BibitemOpen
  \bibfield  {author} {\bibinfo {author} {\bibfnamefont {M.}~\bibnamefont
  {Žnidarič}}, \bibinfo {author} {\bibfnamefont {T.}~\bibnamefont {Prosen}},
  \bibinfo {author} {\bibfnamefont {G.}~\bibnamefont {Benenti}}, \bibinfo
  {author} {\bibfnamefont {G.}~\bibnamefont {Casati}}, \ and\ \bibinfo {author}
  {\bibfnamefont {D.}~\bibnamefont {Rossini}},\ }\href@noop {} {\bibfield
  {journal} {\bibinfo  {journal} {Phys. Rev. E}\ }\textbf {\bibinfo {volume}
  {81}} (\bibinfo {year} {2010})}\BibitemShut {NoStop}%
\bibitem [{\citenamefont {Rebentrost}\ \emph {et~al.}(2009)\citenamefont
  {Rebentrost}, \citenamefont {Mohseni}, \citenamefont {Kassal}, \citenamefont
  {Lloyd},\ and\ \citenamefont {Aspuru-Guzik}}]{EnvironmentAssistedQTransport}%
  \BibitemOpen
  \bibfield  {author} {\bibinfo {author} {\bibfnamefont {P.}~\bibnamefont
  {Rebentrost}}, \bibinfo {author} {\bibfnamefont {M.}~\bibnamefont {Mohseni}},
  \bibinfo {author} {\bibfnamefont {I.}~\bibnamefont {Kassal}}, \bibinfo
  {author} {\bibfnamefont {S.}~\bibnamefont {Lloyd}}, \ and\ \bibinfo {author}
  {\bibfnamefont {A.}~\bibnamefont {Aspuru-Guzik}},\ }\href@noop {} {\bibfield
  {journal} {\bibinfo  {journal} {New Journal of Physics}\ }\textbf {\bibinfo
  {volume} {11}} (\bibinfo {year} {2009})}\BibitemShut {NoStop}%
\bibitem [{\citenamefont {Vedral}(2002)}]{RelativeEntropyVedral}%
  \BibitemOpen
  \bibfield  {author} {\bibinfo {author} {\bibfnamefont {V.}~\bibnamefont
  {Vedral}},\ }\href@noop {} {\bibfield  {journal} {\bibinfo  {journal} {Rev.
  Mod. Phys.}\ }\textbf {\bibinfo {volume} {74}} (\bibinfo {year}
  {2002})}\BibitemShut {NoStop}%
\bibitem [{For()}]{FortranCode}%
  \BibitemOpen
  \href@noop {} {}\bibinfo {note} {The source code of the program used to solve
  the differential equation can be found on:
  \url{https://github.com/EKapetano/LindbladSolver}}\BibitemShut {NoStop}%
\bibitem [{\citenamefont {Cao}\ and\ \citenamefont
  {Silbey}(2009)}]{OptimizationExcitonTrapping}%
  \BibitemOpen
  \bibfield  {author} {\bibinfo {author} {\bibfnamefont {J.}~\bibnamefont
  {Cao}}\ and\ \bibinfo {author} {\bibfnamefont {R.~J.}\ \bibnamefont
  {Silbey}},\ }\href@noop {} {\bibfield  {journal} {\bibinfo  {journal} {J.
  Phys. Chem. A}\ }\textbf {\bibinfo {volume} {113}},\ \bibinfo {pages}
  {13825–13838} (\bibinfo {year} {2009})}\BibitemShut {NoStop}%
\bibitem [{\citenamefont {Thingna}\ \emph {et~al.}(2016)\citenamefont
  {Thingna}, \citenamefont {Manzano},\ and\ \citenamefont
  {Cao}}]{DynamicalSignaturesMolecularSymmetries}%
  \BibitemOpen
  \bibfield  {author} {\bibinfo {author} {\bibfnamefont {J.}~\bibnamefont
  {Thingna}}, \bibinfo {author} {\bibfnamefont {D.}~\bibnamefont {Manzano}}, \
  and\ \bibinfo {author} {\bibfnamefont {J.}~\bibnamefont {Cao}},\ }\href@noop
  {} {\bibfield  {journal} {\bibinfo  {journal} {Sci. Rep.}\ }\textbf {\bibinfo
  {volume} {6}} (\bibinfo {year} {2016})}\BibitemShut {NoStop}%
\bibitem [{\citenamefont {Solomon}\ \emph {et~al.}(2008)\citenamefont
  {Solomon}, \citenamefont {Andrews}, \citenamefont {Hansen}, \citenamefont
  {Goldsmith}, \citenamefont {Wasielewski}, \citenamefont {Duyne},\ and\
  \citenamefont {Ratner}}]{CoherentMolecularConduction}%
  \BibitemOpen
  \bibfield  {author} {\bibinfo {author} {\bibfnamefont {G.~C.}\ \bibnamefont
  {Solomon}}, \bibinfo {author} {\bibfnamefont {D.~Q.}\ \bibnamefont
  {Andrews}}, \bibinfo {author} {\bibfnamefont {T.}~\bibnamefont {Hansen}},
  \bibinfo {author} {\bibfnamefont {R.~H.}\ \bibnamefont {Goldsmith}}, \bibinfo
  {author} {\bibfnamefont {M.~R.}\ \bibnamefont {Wasielewski}}, \bibinfo
  {author} {\bibfnamefont {R.~P.~V.}\ \bibnamefont {Duyne}}, \ and\ \bibinfo
  {author} {\bibfnamefont {M.~A.}\ \bibnamefont {Ratner}},\ }\href@noop {}
  {\bibfield  {journal} {\bibinfo  {journal} {J. Chem. Phys.}\ }\textbf
  {\bibinfo {volume} {129}} (\bibinfo {year} {2008})}\BibitemShut {NoStop}%
\bibitem [{\citenamefont {Misra}\ and\ \citenamefont
  {Sudarshan}(1977)}]{ZenoParadoxQuantumTheory}%
  \BibitemOpen
  \bibfield  {author} {\bibinfo {author} {\bibfnamefont {B.}~\bibnamefont
  {Misra}}\ and\ \bibinfo {author} {\bibfnamefont {E.~C.~G.}\ \bibnamefont
  {Sudarshan}},\ }\href@noop {} {\bibfield  {journal} {\bibinfo  {journal}
  {Journal of Mathematical Physics}\ }\textbf {\bibinfo {volume} {18}},\
  \bibinfo {pages} {756} (\bibinfo {year} {1977})}\BibitemShut {NoStop}%
\bibitem [{\citenamefont {Linke}\ \emph {et~al.}(2002)\citenamefont {Linke},
  \citenamefont {Humphrey}, \citenamefont {Lindelof}, \citenamefont {Löfgren},
  \citenamefont {Newbury}, \citenamefont {Omling}, \citenamefont {Sushkov},
  \citenamefont {Taylor},\ and\ \citenamefont {Xu}}]{QuantumRatchets}%
  \BibitemOpen
  \bibfield  {author} {\bibinfo {author} {\bibfnamefont {H.}~\bibnamefont
  {Linke}}, \bibinfo {author} {\bibfnamefont {T.}~\bibnamefont {Humphrey}},
  \bibinfo {author} {\bibfnamefont {P.}~\bibnamefont {Lindelof}}, \bibinfo
  {author} {\bibfnamefont {A.}~\bibnamefont {Löfgren}}, \bibinfo {author}
  {\bibfnamefont {R.}~\bibnamefont {Newbury}}, \bibinfo {author} {\bibfnamefont
  {P.}~\bibnamefont {Omling}}, \bibinfo {author} {\bibfnamefont
  {A.}~\bibnamefont {Sushkov}}, \bibinfo {author} {\bibfnamefont
  {R.}~\bibnamefont {Taylor}}, \ and\ \bibinfo {author} {\bibfnamefont
  {H.}~\bibnamefont {Xu}},\ }\href@noop {} {\bibfield  {journal} {\bibinfo
  {journal} {Appl. Phys. A}\ }\textbf {\bibinfo {volume} {75}},\ \bibinfo
  {pages} {237} (\bibinfo {year} {2002})}\BibitemShut {NoStop}%
\bibitem [{\citenamefont {Metzger}(2003)}]{UnimolecularRectifiers}%
  \BibitemOpen
  \bibfield  {author} {\bibinfo {author} {\bibfnamefont {R.~M.}\ \bibnamefont
  {Metzger}},\ }\href@noop {} {\bibfield  {journal} {\bibinfo  {journal} {Chem.
  Rev.}\ }\textbf {\bibinfo {volume} {103}},\ \bibinfo {pages} {3803} (\bibinfo
  {year} {2003})}\BibitemShut {NoStop}%
\bibitem [{\citenamefont {Capozzi}\ \emph {et~al.}(2015)\citenamefont {Capozzi}
  \emph {et~al.}}]{SingleMoleculeDiodesHighRectificationEnvironmentalControl}%
  \BibitemOpen
  \bibfield  {author} {\bibinfo {author} {\bibfnamefont {B.}~\bibnamefont
  {Capozzi}} \emph {et~al.},\ }\href@noop {} {\bibfield  {journal} {\bibinfo
  {journal} {Nature Nanotechnology}\ }\textbf {\bibinfo {volume} {10}},\
  \bibinfo {pages} {522} (\bibinfo {year} {2015})}\BibitemShut {NoStop}%
\end{thebibliography}%

\end{document}